\documentclass{appolb}
\usepackage{graphicx}
\usepackage[utf8]{inputenc}
\usepackage{amsmath}
\usepackage{amsfonts}
\usepackage{amssymb}
\usepackage{dsfont}
\usepackage[hidelinks]{hyperref}
\usepackage[]{units}

\DeclareMathOperator{\diag}{diag}

\newcommand{\gtapprox}{\raisebox{-0.5ex}{$\,\stackrel{>}{\scriptstyle\sim}\,$}}

\begin{document}



\title{Refined lattice/model investigation of $u d \bar{b} \bar{b}$ tetraquark candidates with heavy spin effects taken into account\thanks{Presented at Excited QCD 2015: Tatransk\'a Lomnica, Slovakia.}}

\author{Jonas Scheunert$^{(1)}$, Pedro Bicudo$^{(2)}$, Annabelle Uenver$^{(1)}$, Marc Wagner$^{(1)}$
\address{$^{(1)}$~Institut f\"ur Theoretische Physik \\
Goethe-Universit\"at Frankfurt am Main \\
Max-von-Laue-Stra{\ss}e 1, D-60438 Frankfurt am Main, Germany}
\address{$^{(2)}$~CFTP, Dep.\ F\'{\i}sica, Instituto Superior T\'ecnico \\ Universidade de Lisboa \\ Av.\ Rovisco Pais, 1049-001 Lisboa, Portugal}
}

\maketitle

\begin{abstract}
We investigate four-quark systems consisting of two heavy anti-bottom quarks and two light up/down quarks. We propose to solve a coupled Schr\"odinger equation for the anti-bottom-anti-bottom separation using potentials computed via lattice QCD in the limit of static anti-bottom quarks. This coupled Schr\"odinger equation allows to incorporate effects due to the heavy anti-bottom spins. First exploratory numerical tests are discussed.
\end{abstract}

\PACS{12.38.Gc, 13.75.Lb, 14.40.Rt, 14.65.Fy.}


\section{Introduction}

In recent papers \cite{Wagner:2010ad,Wagner:2011ev,Bicudo:2012qt,Wagenbach:2014oxa,Bicudo:2015vta} we have studied heavy tetraquark candidates combining lattice QCD and quark model techniques proceeding in two steps. First we have computed potentials of two static antiquarks $\bar{Q} \bar{Q}$ in the presence of two quarks of finite mass $q q$ ($q \in \{ u,d \}$ throughout this work) using lattice QCD \cite{Wagner:2010ad,Wagner:2011ev} (such potentials have also been computed by other groups, cf.\ e.g.\ \cite{Stewart:1998hk,Michael:1999nq,Cook:2002am,Doi:2006kx,Detmold:2007wk,Bali:2010xa,Brown:2012tm}). The static approximation is expected to be a rather good approximation for $\bar{Q} \bar{Q} = \bar{b} \bar{b}$ and allows for a comparably easy computation of the potentials. For larger $\bar{Q} \bar{Q}$ separations some of these potentials can be interpreted as potentials of two $B$ and/or $B^\ast$ mesons, which are degenerate in the static limit. In a second step we have inserted these potentials into the Schr\"odinger equation for the relative coordinate of the two $B/B^\ast$ mesons. We have then checked, whether they are sufficiently attractive to host bound states, which would indicate stable $q q \bar{b} \bar{b}$ tetraquarks. For a specific potential (isospin $I = 0$, light quark spin $j = 0$) we have found a bound state with confidence level $\approx 2 \sigma$ and binding energy $\approx 90 \, \textrm{MeV}$, while there seems to be no bound state in any of the other channels \cite{Bicudo:2012qt,Bicudo:2015vta}.

In the static limit effects due to the spin of $\bar{b}$ quarks are neglected, e.g.\ there is no mass difference of $B$ and $B^\ast$ for infinitely heavy $\bar{b}$ quarks. These effects, however, could be of the same order as the $\approx 90 \, \textrm{MeV}$ binding energy of the tetraquark predicted in \cite{Bicudo:2012qt,Bicudo:2015vta}, as can e.g.\ be estimated from the mass difference $m_{B^\ast} - m_B \approx 50 \, \textrm{MeV}$. The goal of this work is to take the heavy $\bar{b}$ spins into account, in particular to estimate their effect on the binding energy of the above mentioned $(I = 0 , j = 0)$ $q q \bar{b} \bar{b}$ tetraquark.


\section{Incorporating heavy $\bar{b} \bar{b}$ spin effects}


\subsection{\label{SEC001}Relating $q q \bar{Q} \bar{Q}$ potential and $B^{(\ast)} B^{(\ast)}$ creation operators}

Due to static quark symmetries it is essential to couple the light spin indices and the static spin indices separately, when defining $q q \bar{Q} \bar{Q}$ potential creation operators. To interpret the meson-meson structure generated by such operators, one needs to express them in terms of static-light bilinears. We do this by using the Fierz identity
\begin{equation}
\label{eq:fierz_identity}
	\begin{split}
	& L_{A B} S_{C D} \Big(\bar{Q}_C(\vec{r}_1) q_A^{(1)}(\vec{r}_1)\Big) \Big(\bar{Q}_D(\vec{r}_2) q_B^{(2)}(\vec{r}_2)\Big) \\
	& \quad = \frac{1}{16} \Tr\Big(\Gamma_b S^T \Gamma_a^T L\Big) \Big(\bar{Q}(\vec{r}_1) \Gamma^a q^{(1)}(\vec{r}_1)\Big) \Big(\bar{Q}(\vec{r}_2) \Gamma^b q^{(2)}(\vec{r}_2)\Big) ,
	\end{split}
\end{equation}
where $A,B,C,D$ denote spin indices, \\ $\Gamma^a \in \{ \gamma_5 , \gamma_0 \gamma_5 , \mathds{1} , \gamma_0 , \gamma_j , \gamma_0 \gamma_j , \gamma_j \gamma_5  , \gamma_0 \gamma_j \gamma_5 \}$ and $\Gamma_a$ is the inverse of $\Gamma^a \quad$\footnote{Similar techniques have recently been applied to relate meson-meson and diquark-antidiquark creation operators \cite{Padmanath:2015era}.}. The left hand side of this equation has the structure of a $q q \bar{Q} \bar{Q}$ potential creation operator (cf.\ e.g.\ \cite{Bicudo:2015vta}, eq.\ (6)), while the right-hand side allows to read off, which linear combination of $B$ meson pairs it excites.

In the following we are interested in those matrices $L$ and $S$ generating $B$ and/or $B^\ast$ mesons ($\bar{Q} (\mathds{1} + \gamma_0) \gamma_5 q$ and $\bar{Q} (\mathds{1} + \gamma_0) \gamma_j q$, respectively). After some linear algebra one finds that there are 16 such combinations, $L , \mathcal{C} S^T \mathcal{C} \in \{ \mathcal{C} (\mathds{1} + \gamma_0) \gamma_5 , \mathcal{C} (\mathds{1} + \gamma_0) \gamma_j \}$ ($\mathcal{C}$ denotes the charge conjugation matrix). The $q q \bar{Q} \bar{Q}$ potentials, which have been computed in the static limit, depend only on the light spin coupling $L$, but not on the heavy spin coupling $S$. There are two different potentials, (1)~$V_5(r)$ (corresponding to $L = \mathcal{C} (\mathds{1} + \gamma_0)\gamma_5$), attractive for isospin $I = 0$, repulsive for isospin $I = 1$, and (2)~$V_j(r)$ (corresponding to $L = \mathcal{C} (\mathds{1} + \gamma_0) \gamma_j$), repulsive for isospin $I = 0$, attractive for isospin $I = 1$, where $r = |\vec{r}_1 - \vec{r}_2|$.

Note that it is not possible to choose $S$ and $L$ in a way that exclusively $B$ mesons appear on the right hand side of eq.\ (\ref{eq:fierz_identity}). One always finds linear combinations of $B$ and $B^\ast$ mesons, e.g.\ for $L = \mathcal{C} S^T \mathcal{C} = \mathcal{C} (\mathds{1} + \gamma_0) \gamma_5$ the right hand side of eq.\ (\ref{eq:fierz_identity}) is proportional to $B(\vec{r}_1) B(\vec{r}_2) + B_x^\ast(\vec{r}_1) B_x^\ast(\vec{r}_2) + B_y^\ast(\vec{r}_1) B_y^\ast(\vec{r}_2) + B_z^\ast(\vec{r}_1) B_z^\ast(\vec{r}_2)$ (the indices $x,y,z$ denote the spin orientation of $B^\ast$). Taking this mixing of $B$ and $B^\ast$ mesons into account, which differ in mass by $\approx 50 \, \textrm{MeV}$, is the goal of this work, as already mentioned in the introduction.


\subsection{The coupled channel Schr\"odinger equation}

We study a coupled channel Schr\"odinger equation
\begin{equation}
\label{EQN001} H \Psi(\vec{r}_1,\vec{r}_2) = E \Psi(\vec{r}_1,\vec{r}_2) ,
\end{equation}
where the Hamiltonian $H$ acts on a 16-component wave function $\Psi$. The components of $\Psi$ correspond to the 16 possibilities to combine \\ $(B(\vec{r}_1) , B_x^\ast(\vec{r}_1) , B_y^\ast(\vec{r}_1) , B_z^\ast(\vec{r}_1))$ and $(B(\vec{r}_2) , B_x^\ast(\vec{r}_2) , B_y^\ast(\vec{r}_2) , B_z^\ast(\vec{r}_2))$, i.e.\ the first component corresponds to $B(\vec{r}_1) B(\vec{r}_2)$, the second to $B(\vec{r}_1) B_x^\ast(\vec{r}_2)$, etc.

The Hamiltonian can be split in a free and an interacting part, $H = H_0 + H_\textrm{int}$. The free part is given by
\begin{equation}
H_0 = M \otimes \mathds{1} + \mathds{1} \otimes M + \frac{\vec{p}_1^2}{2} \Big(M \otimes \mathds{1}\Big)^{-1} + \frac{\vec{p}_2^2}{2} \Big(\mathds{1} \otimes M\Big)^{-1}
\end{equation}
with $M = \diag(m_B , m_{B^\ast} , m_{B^\ast} , m_{B^\ast})$. The interacting part can be written according to
\begin{equation}
H_\textrm{int} = T^{-1} V(r) T ,
\end{equation}
where
\begin{equation}
V(r) = \diag\Big(\underbrace{V_5(r) , \ldots V_5(r)}_{4 \times} , \underbrace{V_j(r) , \ldots V_j(r)}_{12 \times}\Big)
\end{equation}
and $T$ is a $16 \times 16$ matrix relating the 16 choices for $L,S$ (cf.\ section~\ref{SEC001} and eq.\ (\ref{eq:fierz_identity})) to the 16 components of $\Psi$ (the entries of $T$ can be computed using the Fierz identity (\ref{eq:fierz_identity})).


\section{\label{SEC003}Numerical solution of the coupled channel Schr\"odinger equation}

Rotational symmetry allows to bring the coupled channel Schr\"odinger equation (\ref{EQN001}) to block diagonal form, i.e.\ to split it into independent simpler equations corresponding to definite total spin $J$ and isospin $I \quad$\footnote{$J$ and $I$ are related, because quarks are fermions and have to obey the Pauli principle (cf.\ \cite{Bicudo:2015vta} for a detailed discussion of quantum numbers of $q q \bar{b} \bar{b}$ tetraquarks).}:
\begin{itemize}
\item a single $2 \times 2$ coupled channel equation:
\\ $J=0$, $I=1$, meson pairs $B(\vec{r}_1) B(\vec{r}_2)$ and $B^\ast(\vec{r}_1) B^\ast(\vec{r}_2)$;

\item three identical $1 \times 1$, i.e.\ uncoupled equations:
\\ $J=1$, $I=1$, meson pairs $B(\vec{r}_1) B^\ast(\vec{r}_2)$ and $B^\ast(\vec{r}_1) B(\vec{r}_2)$;

\item three identical $2 \times 2$ coupled channel equations:
\\ $J=1$, $I=0$, meson pairs $B(\vec{r}_1) B^\ast(\vec{r}_2)$, $B^\ast(\vec{r}_1) B(\vec{r}_2)$ and $B^\ast(\vec{r}_1) B^\ast(\vec{r}_2)$;

\item five identical $1 \times 1$, i.e.\ uncoupled equations:
\\ $J=2$, $I=1$, meson pairs $B^\ast(\vec{r}_1) B^\ast(\vec{r}_2)$.
\end{itemize}

For the remainder of this section we focus on the $J = 0$ coupled channel equation, where
\begin{align}
	\label{eq:H_0,2x2}	
	& H_{0,J=0} =
        \begin{pmatrix}
	2 m_B & 0 \\
	0    & 2 m_{B^\ast}
	\end{pmatrix}
        + \bigg(\frac{\vec{p}_1^2}{2} + \frac{\vec{p}_2^2}{2}\bigg)
	\begin{pmatrix}
	1 / m_B & 0 \\
	0    & 1 / m_{B^{\ast}}
	\end{pmatrix}, \\
	& H_{\textrm{int},J=0} =
        \begin{pmatrix}
	(1/4) (V_5(r) + 3 V_j(r)) & (\sqrt{3}/4) (V_5(r) - V_j(r)) \\
	(\sqrt{3}/4) (V_5(r) - V_j(r)) & (1/4) (3 V_5(r) + V_j(r))
	\end{pmatrix} .
\end{align}
Introducing center of mass and relative coordinates the partial differential equation in $\vec{r}_1$ and $\vec{r}_2$ can analytically be reduced to an ordinary differential equation for $r$,
\begin{equation}
\label{EQN002} \left(
\begin{pmatrix}
2 m_B - \frac{1}{m_B} \frac{\mathrm{d}^2}{\mathrm{d}r^2} & 0 \\
0    & 2 m_{B^\ast} - \frac{1}{m_{B^\ast}} \frac{\mathrm{d}^2}{\mathrm{d}r^2}
\end{pmatrix}
+ H_{\textrm{int},J=0}
\right) \chi(r) = E \chi(r) ,
\end{equation}
where the first component of $\chi$ represents a $B(\vec{r}_1) B(\vec{r}_2)$ pair and the second component a $B^\ast(\vec{r}_1) B^\ast(\vec{r}_2)$ pair. Following standard textbooks on quantum mechanics one can show that the radial wave function of an $s$ wave bound state is subject to the boundary conditions
\begin{equation}
\label{EQN003} \chi(r) \sim \begin{pmatrix} A r \\ B r \end{pmatrix} \textrm{ as } r \rightarrow 0
\quad , \quad
\lim_{r \rightarrow \infty} \chi(r) =  \begin{pmatrix} 0 \\ 0 \end{pmatrix}
\end{equation}
with $A , B \in \mathds{R}$.

First exploratory numerical tests have been performed with $I = 0$ potentials, i.e.\ with an attractive $V_5$ and a weakly repulsive $V_j \quad$\footnote{Even though this $(J=0 , I=0)$ channel is excluded by the Pauli principle, it is conceptually interesting to compare numerical results with existing results from \cite{Bicudo:2012qt,Bicudo:2015vta}, where heavy spin effects have not been taken into account.}.  We integrate eq.\ (\ref{EQN002}) using the Runge-Kutta-Fehlberg method starting with the linear asymptotic behavior (\ref{EQN003}) at tiny $r = \varepsilon > 0$ to $r = r_\textrm{max}$ with sufficiently large $r_\textrm{max} \gtapprox 10 \, \textrm{fm}$. This integration is iterated many times as part of a standard shooting procedure to find parameters $A/B$ and $E$ such that also $\chi_1(r_\textrm{max}) = \chi_1(r_\textrm{max}) = 0$ is fulfilled.

\begin{figure}[htb]
\centerline{\includegraphics[width=8.0cm]{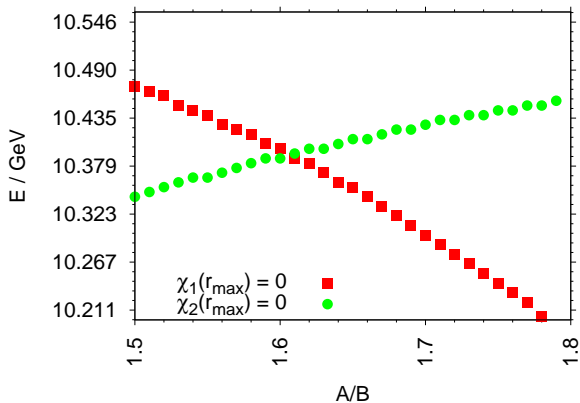}}
\caption{\label{fig:plot}Isolines $\chi_1(r_\textrm{max}) = 0$ (red) and $\chi_2(r_\textrm{max}) = 0$ (green) in the $A/B$-$E$ plane (for unphysically strong attractive potential $V_5$).}
\end{figure}

In Fig.~\ref{fig:plot} we show results obtained with an unphysically strong attractive potential $V_5$ (roughly a factor $1.5$ stronger than the lattice QCD result for $V_5$). The intersection of the red line and the green line at $E \approx 10.4 \, \textrm{GeV}$ corresponds to $\chi_1(r_\textrm{max}) = \chi_1(r_\textrm{max}) = 0$, i.e.\ represents an energy eigenstate. Since $E < 2 m_B \approx 10.6 \, \textrm{MeV}$ ($2 m_B$ is the upper boundary of the plot), this eigenstate is a bound four quark state.

When repeating these calculations with crude fits to the lattice QCD results for $V_5$ and $V_j$, the situation is less clear, i.e.\ $E \approx 2 m_B$. A more careful analysis and treatment of statistical and systematic errors similar to what has been done in \cite{Bicudo:2015vta} is needed, to confirm or rule out a bound state. In any case, one can conclude that the heavy $\bar{b} \bar{b}$ spins counteract four-quark binding.


\section{Outlook}

Most interesting will, of course, be an investigation of the physical channels listed at the beginning of section~\ref{SEC003}, in particular the $(J=1 , I=0)$ channel, which has a stronger attractive potential than the $I=1$ channels. We are currently in the process of studying corresponding Schr\"odinger equations for all these channels.


\section*{Acknowledgments}

J.S.\ thanks the organizers of ``Excited QCD 2015'' for the opportunity to give this talk. P.B.\ thanks IFT for hospitality and CFTP, grant FCT \\ UID/FIS/00777/2013, for support. M.W.\ acknowledges support by the Emmy Noether Programme of the DFG (German Research Foundation), grant WA 3000/1-1. This work was supported in part by the Helmholtz International Center for FAIR within the framework of the LOEWE program launched by the State of Hesse. Calculations on the LOEWE-CSC high-performance computer of Johann Wolfgang Goethe-University Frankfurt am Main were conducted for this research. We would like to thank HPC-Hessen, funded by the State Ministry of Higher Education, Research and the Arts, for programming advice.



\end{document}